\documentclass[aps,prd,floatfix,nofootinbib]{revtex4} 
\usepackage{graphics}
\usepackage{graphicx}

\usepackage{latexsym}
\usepackage[cp866]{inputenc}
\usepackage[english]{babel}
\input epsf

\begin{document}

\title{
Qualitative explanation of the data on the decays $D^0\to
a_0(980)^\pm\pi^\mp$ and $D^+\to a_0(980)^{+(0)}\pi^{0(+)}$ in the
four-quark model of the $a_0(980)$ resonance}
\author{N. N. Achasov\,\footnote{achasov@math.nsc.ru}
and G. N. Shestakov\,\footnote{shestako@math.nsc.ru}}
\affiliation{\vspace{0.2cm} Laboratory of Theoretical Physics, S. L.
Sobolev Institute for Mathematics, 630090, Novosibirsk, Russia}

\begin{abstract}

It is shown that the values of the ratios $\mathcal{B}(D^0\to
a_0(980)^+\pi^-)/\mathcal{B} (D^0\to a_0(980)^-\pi^+ )$ and
$\mathcal{B}(D^+\to a_0(980)^+\pi^0)/\mathcal{B} (D^+\to a_0(980)^0
\pi^+)$, recently measured by the BESIII Collaboration, are
naturally explained in the four-quark model of the $a_0(980) $
resonance.

\end{abstract}

\maketitle
\section{\boldmath Introduction}

Phenomenological studies of two-particle and quasi-two-particle
hadronic decays of $D$-mesons using the quark-diagram scheme have a
long and seminal history, see, for example, Refs. \cite{CC87,Ro99,
CC10,CC10a,CCK16,CCZ22,Ab24}. Recently, the BESIII Collaboration
\cite{Ab24} presented the first amplitude analysis of the
Cabibbo-suppressed decays $D^0\to\pi^+\pi^-\eta$ and $D^+\to\pi^+
\pi^0\eta $, as a result of which, for the first time, information
about the contributions from the $a_0(980)\pi$ intermediate states
into these decays was obtained with high statistical significances.
In particular, for the ratios $\mathcal{B}(D^0\to a_0(980)^+\pi^-)
/\mathcal{B} (D^0\to a_0(980)^-\pi^+ )$ and $\mathcal{B}(D^+\to a_0
(980)^+\pi^0)/\mathcal{B} (D^+\to a_0(980)^0\pi^+)$ were obtained
the following data \cite{Ab24}:
\begin{eqnarray}\label{Eq1}
r_{+/-}=\frac{\mathcal{B}(D^0\to a_0(980)^+\pi^-)}{\mathcal{B}
(D^0\to a_0(980)^-\pi^+)}=7.5^{+2.5}_{-0.8}\pm1.7 \quad \mbox{and}
\quad r_{+/0}=\frac{\mathcal{B}(D^+\to a_0(980)^+\pi^0)}{\mathcal{B}
(D^+\to a_0(980)^0\pi^+)}=2.6\pm0.6\pm0.3.
\end{eqnarray}
In the quark-diagram scheme, it was expected that the value of the
ratio $r_{+/-}$ would be less than 0.05 \cite{CCZ22,Ab24,FN1}. In
this note, we propose a simple qualitative explanation of the values
of the ratios $r_{+/-}$ and $r_{+/0}$ in the four-quark model for
the $a_0(980) $-resonance \cite{Ja77,Ac80a,Ac89,Ac98, Ac11}, see
also Ref. \cite{SW13}. In this model, $a_0(980)^{(\pm,0)}$ mesons
are states with symbolic quark structures of the form:
$a_0(980)^+=u\bar ds\bar s$, $a_0(980)^0= s\bar s(u\bar u-d\bar
d)/\sqrt{2}$, $a_0(980)^-=d\bar us\bar s$ \cite{Ja77}.

\section{\boldmath The quark-diagram scheme}

The formation of the four-quark $a_0(980)$ meson in a pair with the
$\pi$ meson in $D$ decays  can occur as a result of various quark
fluctuations. The simplest quark diagrams giving examples of such
fluctuations are shown in Figs. 1--4. The first diagram in Fig. 1
corresponds to the situation when after the $c$ quark decay in the
$D^0(c\bar u)$ meson, $c\bar u\to(u\bar ss)\bar u$, the direct
formation of the four-quark $a_0(980)^+$ meson and the $\pi^-$ meson
can occur owing to  the vacuum $\bar dd$ fluctuation: $D^0\to [c\bar
u\to(u\bar ss)\bar u\to(u\bar ss\bar d)(d\bar u)]\to a_0(980)^+
\pi^-$. This diagram is the main element in our scheme (it also
appears in Figs. 3 and 4). The decay of $D^0\to a_0(980)^+\pi^-$
with the formation of the four-quark $a_0(980)$ meson is also
possible due to the second diagram in Fig. 1. Here the production of
the $a_0(980)^+\pi^-$ system occurs via the $d\bar d$ intermediate
state as a result of the double vacuum fluctuation $(\bar uu)(\bar
ss)$. The decay of $D^0\to a_0(980)^-\pi^+$ is described by the
diagrams in Fig. 2. The diagram in Fig. 3 and diagrams in Fig. 4
correspond to the decays $D^+\to a_0(980)^+ \pi^0$ and $D^+\to
a_0(980)^0\pi^+$, respectively. Naturally, there are no diagrams
with $W^+$ exchange between the valence quarks in the $D^+$ meson,
and contributions from the annihilation diagrams $D^+\to W^+\to
a_0(980)^{+(0 )}\pi^{0(+)}$ are suppressed since G-parity  is
violated in such transitions \cite{Ac17}. Let us write the ratios
$r_{+/-}$ and $r_{+/0}$ in terms of three amplitudes $a$, $b$ and
$e$ corresponding to the diagrams in Fig. 1--4 (see figure
captions):
\begin{eqnarray}\label{Eq2}
r_{+/-}=\frac{|a+e|^2}{|b+e|^2} \quad \mbox{and} \quad
r_{+/0}=\frac{|a|^2}{|a-b|^2}. \end{eqnarray} Let us restrict
ourselves to a simplest possible solution of these relations. We
neglect the amplitude $e$ due to the double vacuum fluctuation and
assume that the amplitudes $a$ and $b$ are real and identical in
sign. Then Eq. (\ref{Eq2}) leads to the following relation between
$r_{+/-}$ and $r_{+/0}$:
\begin{eqnarray}\label{Eq3}
r_{+/0}=\frac{r_{+/-}}{|\sqrt{r_{+/-}}-1|^2}.
\end{eqnarray}
We put $r_{+/-}=7.5$ [see. Eq. (\ref{Eq1})]. Then from Eq.
(\ref{Eq3}) we find $r_{+/0}\approx2.48$. And vice versa, assuming
$r_{+/0}=2.6$ [see Eq. (\ref{Eq1})], from Eq. (\ref{Eq3}) we get
$r_{+/-}\approx6.93$. How $r_{+/0}$ changes depending on $r_{+/-}$
is shown in Fig. 5. Thus, with the assumptions made, the structure
of the mechanism of the four-quark $a_0(980)$ meson production in
the $D\to a_0(980)\pi$ decays [see Figs. 1--4 and Eq. (\ref{Eq2})]
allows us to qualitatively understand the BESIII data \cite{Ab24}
for $r_{+/-}$ and $r_{+/0}$ and even reasonably  describe them by
using Eq. (\ref{Eq3}). Note that the previously obtained evidences
in favor of the four-quark nature of the $a_0(980)$ state
\cite{Ja77,Ac80a, Ac89,Ac98,Ac11} are based on the analyzes of
experimental results
and confirmed by experiments. \\[0.2cm]

\begin{figure}  [!ht] 
\begin{center}\includegraphics[width=11.4cm]{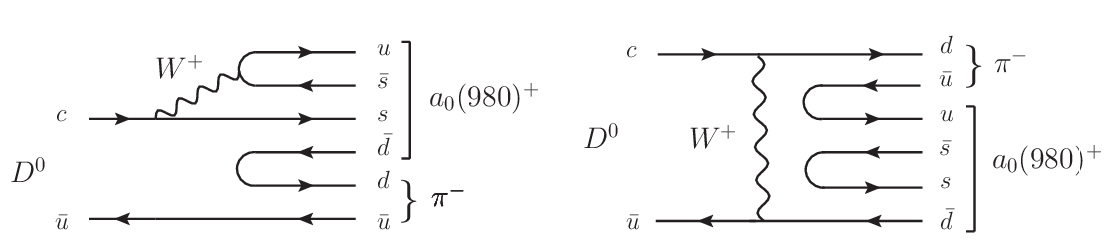}
\caption{\label{Fig1} Diagrams for $D^0\to a_0(980)^+\pi^-$. The
amplitudes of the first and second diagrams are labeled as $a$ and
$e$, respectively.}
\end{center}\end{figure}
\begin{figure}  [!ht] 
\begin{center}\includegraphics[width=11.4cm]{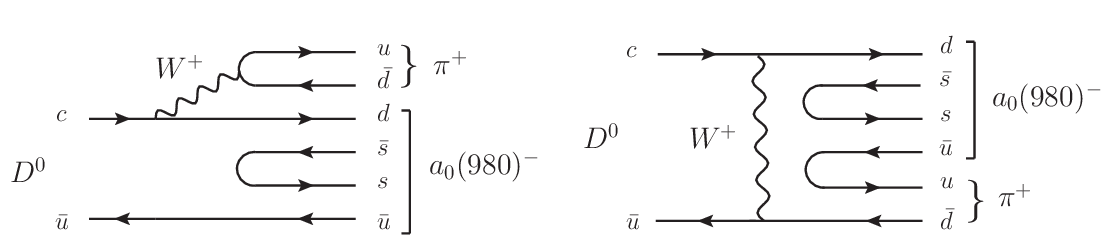}
\caption{\label{Fig2} Diagrams for $D^0\to a_0(980)^-\pi^+$. The
amplitude of the first diagram is labeled as $b$. The amplitude of
the second diagram is equal to $e$ (see Fig. 1 caption).}
\end{center}\end{figure}
\begin{figure}  [!ht] 
\begin{center}\includegraphics[width=5.5cm]{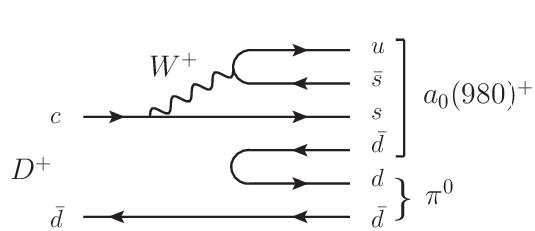}
\caption{\label{Fig3} The diagram for $D^+\to a_0(980)^+\pi^0$. The
amplitude of this diagram is equal to $-a/\sqrt{2}$ (see Fig. 1
caption).}\end{center}\end{figure}
\begin{figure}  [!ht] 
\begin{center}\includegraphics[width=11.4cm]{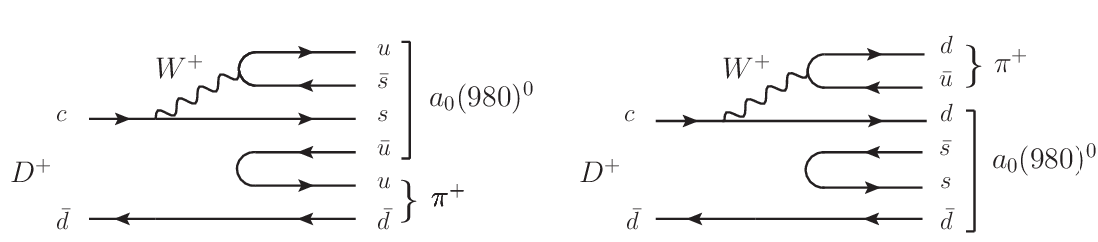}
\caption{\label{Fig4} Diagrams for $D^+\to a_0(980)^0\pi^+$. The
amplitudes of the first and second diagrams are equal to $a/\sqrt{
2}$ and $-b/\sqrt{2}$, respectively (see captions for Figs. 1 and
2).}\end{center}\end{figure}
\begin{figure}  [!ht] 
\begin{center}\includegraphics[width=5.5cm]{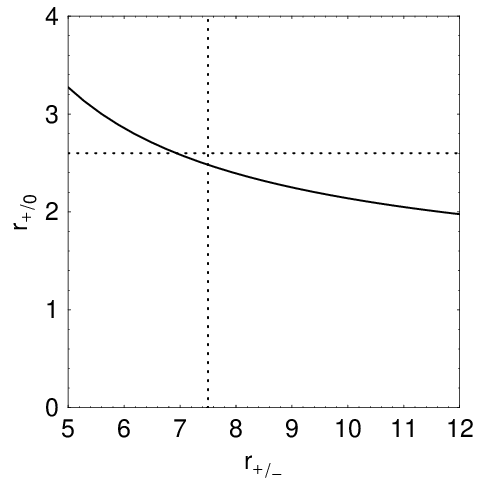}
\caption{\label{Fig4} The solid curve shows $r_{+/0}$ as a function
of $r_{+/-}$ [see Eq. (\ref{Eq3})] within the uncertainties of
$r_{+/-}$ indicated in Eq. (\ref{Eq1}). The the dotted vertical and
horizontal lines mark the central values of $r_{+/-}$ and $r_{+/0}$
in Eq. (\ref{Eq1}), respectively.}\end{center}\end{figure}

\newpage

Let us now provide a few additional comments. Certainly, the BESIII
data \cite{Ab24} demonstrate the importance of taking into account
the amplitudes of seed multiquark fluctuations along with the
simplest seed quark amplitudes $T$, $C$, $E$ and $A$
\cite{Ro99,CC10}. Of course, our assumption about the seed
amplitudes $a$ and $b$ is rather rough. These amplitudes must be
dressed by $a_0(980)\pi$ interactions in the final state and acquire
phases in accordance with their decomposition into components with a
defined isospin. It is remarkable that more than 10 years ago an
example of the $S$-wave interaction amplitude between the $\pi$ and
$a_0(980)$ mesons was constructed, taking into account the
$a_0(980)$ coupling to the $K\bar K$-channels \cite{AOR10}. The
problem of taking into account the $a_0(980)\pi$ final state
interactions requires further careful study.

Of course, the quark processes in Figs. 1--4 can be taken as an
initial start out of which to build the $a_0(980)$ resonance due to
coupled-channel interactions involving the $K\bar K$ and $\pi\eta$
channels. For instance, this was similarly done in Ref. \cite{MO01}
for $J/\psi\to\phi\pi\pi(K\bar K)$ decays, where final state
interactions are taken into account as rescattering effects in the
system of the two pseudoscalar mesons.

Here, it would also be appropriate to note the molecular $K\bar K$
model for the $a_0(980)$ state (see for review Refs.
\cite{Ac98,Ac11}. In these reviews, it was noted that in a number of
cases the four-quark and molecular models for the $a_0(980)$- and
$f_0(980)$-mesons lead to similar predictions, but there are also a
number of significant differences, for example, in predictions for
their two-photon widths, and these differences do not favor the
molecular model \cite{Ac98,Ac11}. As for the considered case of the
$D\to a_0(980)\pi$ decays, it is difficult to indicate any
differences between these models in the language of the symbolic
quark amplitudes.

\vspace*{0.3cm}

\begin{center} {\bf ACKNOWLEDGMENTS} \end{center}

The work was carried out within the framework of the state contract
of the Sobolev Institute of Mathematics, Project No. FWNF-2022-0021.




\begin{thebibliography}{99}
\bibitem{FN1} Difficulties in describing the $D\to SP$ decays (where $S$ and $P$ denote the light scalar and pseudoscalar mesons, respectively) %
in the quark-diagram scheme were discussed, for example, in Refs. \cite{CC10a,CCZ22,Ab24}. %
\end{thebibliography}
\end{document}